\documentclass[aps,pra,twocolumn,nofootinbib,superscriptaddress]{revtex4-2}
\usepackage{amsmath,amssymb,bm,graphicx,xcolor}
\usepackage[colorlinks=true,linkcolor=blue,citecolor=blue,urlcolor=blue]{hyperref}
\usepackage{braket}
\usepackage{float}
 \usepackage{orcidlink}
\begin{document}

\title{Harnessing Environmental Memory with Reinforcement Learning in Open Quantum Systems}
\author{Safae Gaidi \orcidlink{0009-0003-9635-0274}}
\affiliation{LPHE-MS, Faculty of Sciences, Mohammed V University in Rabat, Rabat, Morocco}
\author{Abdallah Slaoui \orcidlink{0000-0002-5284-3240}}\affiliation{LPHE-MS, Faculty of Sciences, Mohammed V University in Rabat, Rabat, Morocco}\affiliation{CPM, Faculty of Sciences, Mohammed V University in Rabat, Rabat, Morocco.}
\author{Mohammed EL Falaki \orcidlink{0000-0001-6835-8090}}\affiliation{LPHE-MS, Faculty of Sciences, Mohammed V University in Rabat, Rabat, Morocco}
\author{Amine Jaouadi \orcidlink{0000-0001-8155-8011}}
\email{ajaouadi@ece.fr}
\affiliation{LyRIDS, ECE Engineering School, OMNES Education, 10 rue Sextius Michel, 75015 Paris - France}

\date{\today}

\begin{abstract}
Non-Markovian quantum dynamics, characterized by information backflow from the environment to the system, has emerged as a potential resource for quantum technologies. A key challenge is therefore to control and enhance such memory effects. In this work, we investigate the use of reinforcement learning (RL) to maximize non-Markovianity in a driven two-level system coupled to a structured reservoir. We compare RL-based control strategies with standard optimal control theory (OCT).
We show that OCT produces localized but relatively weak revivalsin the instantaneous non-Markovianity rate, whereas RL policies generate significantly stronger and better-timed information backflow by synchronizing the system dynamics with favorable memory intervals of the environment. This enhanced exploitation of memory effects leads to a higher total integrated non-Markovianity for RL than for OCT, with SAC achieving the largest overall enhancement and PPO delivering slightly lower but still strongly improved performance with smoother, experimentally attractive pulses. Our results contribute to the emerging view of non-Markovianity as an operational resource and illustrate how RL can serve as a flexible, model-free tool for non-Markovian quantum control.

\end{abstract}

\keywords{
Non-Markovianity; 
Open quantum systems; 
Reinforcement learning; 
Quantum control; 
Information backflow
}

\maketitle

\section{Introduction}

The interaction between a quantum system and its surrounding environment 
leads to decoherence, energy relaxation, and loss of distinguishability 
between quantum states~\cite{Breuer2002,ARivas2012}. In many physical 
platforms from solid-state spins and superconducting qubits to 
nanophotonic structures and molecular systems~\cite{Mangaud2017ChemPhys} the environment is structured or strongly coupled, and its 
correlations persist on timescales comparable to the intrinsic system 
dynamics. In such regimes, the reduced dynamics cannot be described by 
a simple Markovian semigroup and instead exhibits memory effects, 
revivals of coherence, and information backflow, i.e.,signatures of genuinely non-Markovian behavior~\cite{Breuer2009,Rivas2014,deVega2017,Shahri2023,Chruscinski2014,Bylicka2014}.
Over the past decade, non-Markovianity has been recognized as a useful resource rather than merely a nuisance. It can enhance quantum channel capacities and improve the performance of information-processing tasks, for example by correlating memory effects with higher dense coding capacities in non-Markovian regimes ~\cite{Rivas2014,Dakir2025}. It has also been shown that temporal correlations from non-Markovian environments can be consumed to reduce noise beyond the reach of standard dynamical decoupling techniques~\cite{Berk2023}. Moreover, appropriately controlled non-Markovian effects can offer advantages in quantum metrology and sensing, enabling precision beyond the standard quantum limit~\cite{Yang2024,Gaidi2025}. 
\newline
In parallel, significant efforts have been devoted to understanding the microscopic origin of non-Markovian dynamics in driven open quantum systems, where deriving physically consistent time-local master equations remains a highly nontrivial task. Notably, microscopic treatments for driven two-level systems coupled to structured reservoirs have been developed under well-defined approximations \cite{Haikka2010}. 
More generally, non-Markovian information backflow and system–environment memory can qualitatively alter open quantum dynamics with implications for quantum communication, simulation, and sensing protocols \cite{Yang2024}.
From a theoretical standpoint, controlling non-Markovian dynamics is challenging. Memory kernels, time-nonlocal master equations, or time-dependent decay rates such as $\gamma(t)$ in time-local formulations lead to control landscapes that are highly non-convex and history dependent~\cite{Cong2024}. Standard optimal control theory (OCT) methods, which have been remarkably successful in Markovian and closed-system settings ~\cite{Goerz2014, Glaser2015}, have also been explored to control the non-Markovian response under strong driving~\cite{PuthumpallyJoseph2018, Mangaud2018},
but rely on the computation of gradients of a final-time cost functional with respect to control fields. In the presence of memory, these gradients become sensitive to numerical approximations and to the full past trajectory, which can lead to unstable optimization and trapping in local optima~\cite{KochGoerz2022}. This difficulty is amplified when the figure of merit is itself non-linear and temporally non-local, as is the case for most non-Markovianity measures.

In parallel, there has been rapid progress in using machine learning to characterize and exploit non-Markovian dynamics. Data-driven schemes have been proposed to reconstruct memory kernels and effective environments directly from experimental trajectories~\cite{Luchnikov2020,Gupta2021}, and to analyze complex memory effects in many-body systems~\cite{Liu2019,Trivino2024}. These approaches highlight that non-Markovianity can be accessed operationally through repeated interaction and statistical learning, without the need for an exact microscopic model.

Reinforcement learning (RL) naturally fits within this paradigm. In RL, an agent learns a control policy by interacting with the system and maximizing a scalar reward signal~\cite{Sutton1998,FrancoisLavet2018}. RL has been successfully applied to quantum state preparation, gate design, and feedback control in both simulations and experiments~\cite{Bukov2018,Porotti2022,Khalid2023}. In particular, RL can be implemented in a model-free fashion directly on hardware, allowing control strategies to adapt to the specific, possibly non-Markovian noise of the device~\cite{Sivak2021,An2019}. Recent work has begun to combine RL with physics-informed constraints and model-based elements to improve sample efficiency and robustness in open quantum systems~\cite{WangWu2024,Moll2021}.

Very recently, RL has also been explored explicitly in non-Markovian settings. Hybrid schemes integrating model learning and RL have been proposed for spin-boson models with memory~\cite{Neema2024}, RL has been shown to provide an efficient alternative to conventional OCT in non-Markovian molecular control problems~\cite{Jaouadi2024}. At the same time, independent developments in non-Markovian quantum control and reservoir engineering~\cite{Cong2024,Lacroix2024} underscore the importance of designing control protocols that are sensitive to the temporal structure of environmental memory.

In this work, we bring together these threads by investigating how modern RL methods can actively exploit environmental memory in a minimal yet representative open quantum system. We consider a driven two-level system coupled to a Lorentzian reservoir, where the time-dependent decay rate $\gamma(t)$ can become temporarily negative, signaling information backflow \cite{Gaidi2025b,Haarnoja2018,Fosel2018}. Using measure of non-Markovianity as our objective, we design a reward that directly favors episodes of positive trace-distance growth and implement both a Soft Actor-Critic(SAC) and a Proximal Policy Optimization (PPO) agent to learn continuous control fields \cite{Niu2019,Dalgaard2020,Goerz2022}. We benchmark these RL policies against standard OCT with Powell and L-BFGS-B. We show that standard OCT produces a relatively weak and localized enhancement of the instantaneous non-Markovianity, confined to the first available memory window. In contrast, RL policies generate substantially larger information backflow by synchronizing the system dynamics with more favorable memory intervals of the environment. In particular, SAC achieves the highest peak values of the instantaneous non-Markovianity and the largest total integrated non-Markovianity, while PPO reaches slightly lower but still strongly enhanced performance with smoother control profiles.
This improvement does not arise from distributing small revivals over time, but rather from the ability of RL to reshape the dynamics and exploit memory effects more efficiently at the most favorable times. As a result, the integrated non-Markovianity achieved by RL significantly exceeds that obtained with OCT. These findings reinforce the view of non-Markovianity as an operational resource and highlight the potential of reinforcement learning as a flexible, model-free approach to controlling memory effects in open quantum systems.

The remainder of this paper is organized as follows.
In Sec.~II, we introduce the theoretical model of a driven two-level
system coupled to a structured bosonic reservoir and recall the
conditions under which non-Markovian memory effects arise.
In Sec.~III, we define the measure of non-Markovianity based on the
Breuer--Laine--Piilo trace-distance criterion, which serves as the
central figure of merit throughout this work.
Section~IV is devoted to a benchmark study using standard optimal
control theory, where Powell and L-BFGS-B algorithms are employed to
maximize the total non-Markovianity and to analyze the resulting
limitations induced by memory-dependent control landscapes.
In Sec.~V, we introduce the RL framework, including
the environment formulation, the reward design, and the implementation
of the PPO and SAC algorithms for continuous quantum control.
The numerical results and a detailed comparison between optimal control
and reinforcement learning are presented and discussed in Sec.~VI,
highlighting the distinct strategies uncovered by each approach.
Finally, Sec.~VII summarizes our findings and outlines future directions
for exploiting non-Markovian memory effects using data-driven quantum
control methods.

\section{Theoretical Model}

We consider a driven two-level system with Hamiltonian
\begin{equation}
H(t)=\frac{\Delta}{2}\sigma_z + \frac{\Omega(t)}{2}\sigma_x,
\label{eq:1}
\end{equation}
where $\Delta$ is the detuning and $\Omega(t)$ is a controllable driving field. The system interacts with a structured bosonic reservoir, resulting in a time-local master equation \cite{Breuer2002}:
\begin{equation}
\dot{\rho}(t) = -i[H(t),\rho(t)] + \gamma(t)\!\left(
\sigma_-\rho\sigma_+ - \tfrac{1}{2}\{\sigma_+\sigma_-,\rho\}
\right).
\label{eq:master}
\end{equation}

For a Lorentzian spectral density, the decay rate is \cite{Shahri2023,Gaidi2025b}:
\begin{equation}
\gamma(t)=\Re\!\left[
\frac{2\Gamma\lambda\sinh(dt/2)}{d\cosh(dt/2)+\lambda\sinh(dt/2)}
\right],
\label{eq:3}
\end{equation}
where $d=\sqrt{\lambda^2-2\Gamma\lambda}$. When $\Gamma>\lambda/2$, the decay rate becomes negative in certain intervals-a signature of information backflow and non-Markovianity.

The quantity $\gamma(t)$ plays a central role in characterizing memory
effects. In the weak-coupling regime $\Gamma < \lambda/2$, $\gamma(t)$ remains
strictly positive and the dynamics is CP-divisible, corresponding to a
Markovian amplitude-damping process.  
In contrast, for $\Gamma > \lambda/2$, the decay rate exhibits pronounced
oscillations with intervals of negativity. These $\gamma(t)<0$ windows signal
genuine information backflow from the environment into the system and thus
constitute the non-Markovian regime \cite{Breuer2009,Rivas2014,deVega2017}.
Throughout this work, we set $\Gamma = 1$ and $\lambda = 0.5$, which 
produces a single revival while remaining in the non-Markovian regime, 
representing the minimal scenario for studying memory-enhanced optimization.

\section{Non-Markovianity Measure}

We quantify information backflow using the BLP measure \cite{Breuer2009}, based on the trace distance:
\begin{equation}
D(t)=\tfrac{1}{2}\|\rho_1(t)-\rho_2(t)\|.
\label{eq:4}
\end{equation}
An increase in distinguishability,
\begin{equation}
\dot{D}(t)>0,
\label{eq:5}
\end{equation}
indicates information flowing from the environment back into the system.

We define the instantaneous non-Markovianity rate
\begin{equation}
\mathcal{N}_{\rm loc}(t)=\max[0,\dot{D}(t)],
\label{eq:6}
\end{equation}
and the total non-Markovianity over the interval $[0,T]$:
\begin{equation}
\mathcal{N}_{Tot}= \int_0^T \mathcal{N}_{\rm loc}(t)\,dt.
\label{eq:7}
\end{equation}
Following the standard result for amplitude-damping channels, we choose optimal initial states $\rho_1(0)=|1\rangle\langle1|$ and $\rho_2(0)=|0\rangle\langle0|$.

\section{Optimal Control Theory}

We now evaluate whether OCT can enhance non-Markovianity by shaping the driving field $\Omega(t)$. The control pulse is parameterized as piecewise constant with $N_c$ time bins and bounded amplitude. We use two established OCT algorithms: Powell’s derivative-free method, and L-BFGS-B, a limited-memory quasi-Newton algorithm with bound constraints.These algorithms are widely used in quantum control \cite{Goerz2014}.

\subsection{Standard OCT convergence: total non-Markovianity}

\begin{figure}[h]
\centering
\includegraphics[width=\linewidth]{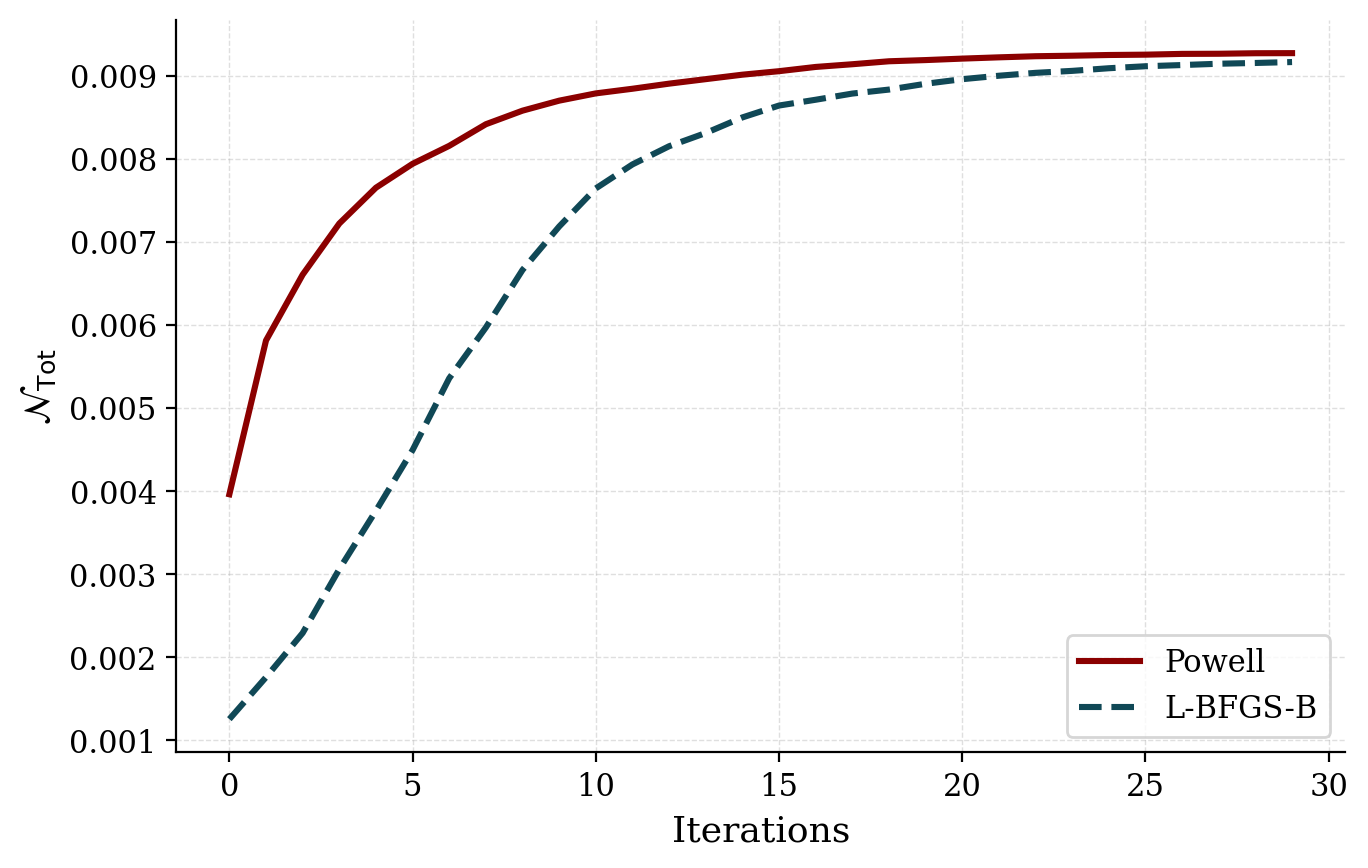}
\caption{Convergence history of the total non-Markovianity $\mathcal{N}{\mathrm{Tot}}$, averaged over 5 independent runs with different random seeds, for Powell (red solid curve) and L-BFGS-B (blue dashed curve). Both algorithms converge to low plateau values, around $\mathcal{N}{\mathrm{Tot}}\simeq 0.009$, indicating that standard OCT provides only a limited enhancement of non-Markovianity and tends to become trapped in local optima of the memory-dependent control landscape.}
\label{fig:total}
\end{figure}

To evaluate the performance of standard OCT, we directly maximize the total non-Markovianity \(\mathcal{N}_{\mathrm{Tot}}\) by optimizing the set of control pulse amplitudes \(\{\Omega_j\}\). Figure~\ref{fig:total} shows the evolution of \(\mathcal{N}_{\mathrm{Tot}}\) as a function of the iteration number for the Powell and L-BFGS-B methods, highlighting their distinct convergence behaviors.

Figure~\ref{fig:total} shows the averaged convergence of the total non-Markovianity over five independent OCT runs with different random initial pulses. Both Powell and L-BFGS-B reach low plateau values, around $(\mathcal{N}_{\mathrm{Tot}}\simeq 0.009)$, showing that the limited standard OCT performance is not a consequence of a single unfavorable initialization.

The corresponding instantaneous non-Markovianity profile shown in Fig.~\ref{fig:oct_local} is obtained from the best pulse selected among the independent OCT runs. This best-case pulse produces a weak and localized revival in the first memory window, around $(t\simeq 4-8)$, with a maximum value $(N_{\mathrm{loc}}^{\max}\approx 0.0065)$. Thus, even the best OCT pulse found across the independent runs only generates a small local backflow peak, consistent with the low accumulated value observed in Fig.~\ref{fig:total}.

This behavior indicates that standard OCT becomes trapped in local optima of the control landscape, which is shaped by non-linear and history-dependent memory effects. Such limitations of gradient-based optimization in non-Markovian regimes have been discussed in previous works \cite{Goerz2014,KochGoerz2022}.

\subsection{OCT effect on instantaneous non-Markovianity}

\begin{figure}[h]
\centering
\includegraphics[width=\linewidth]{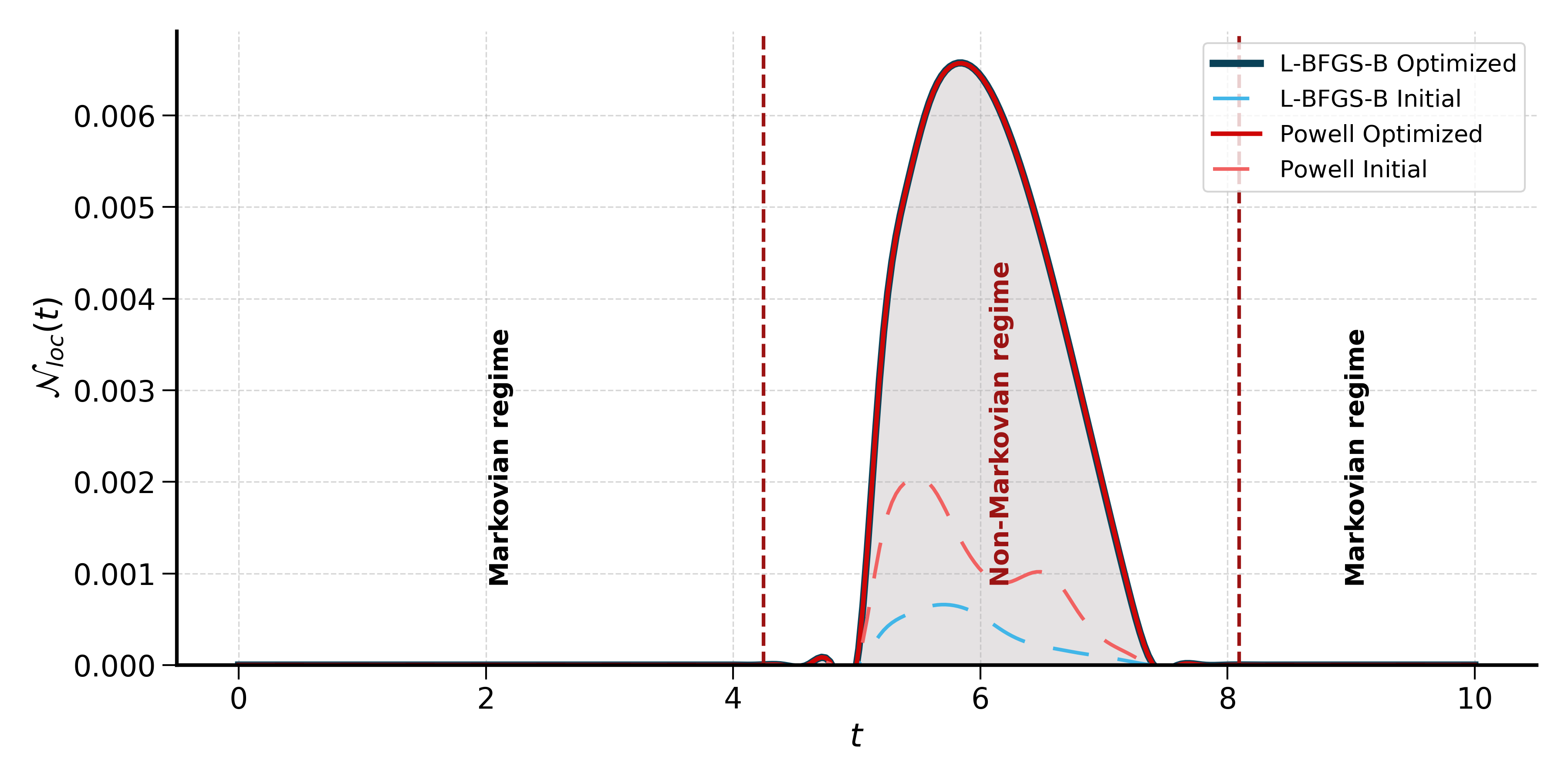}
\caption{Instantaneous non-Markovianity $\mathcal{N}*{\rm loc}(t)$ corresponding to the best pulse found over 5 independent runs, for Powell (red) and L-BFGS-B (blue). Shaded areas indicate the non-Markovian regime where $\mathcal{N}_{\rm loc}(t) > 0$.}
\label{fig:oct_local}
\end{figure}

To analyze how standard OCT reshapes the temporal structure of information backflow, we examine the instantaneous rate $\mathcal{N}_{\rm loc}(t)$. Figure~\ref{fig:oct_local} shows $\mathcal{N}_{\rm loc}(t)$ for the best pulse selected among five independent optimization runs, compared to the uncontrolled case.

The optimized pulse leads to a visible but limited enhancement of the backflow, with a dominant revival occurring in the first memory window around $t \simeq 4-8$. The maximal value remains small, $N_{\mathrm{loc}}^{\max} \approx 0.0065$, indicating that standard OCT only weakly amplifies the instantaneous non-Markovianity.

This behavior reflects the sensitivity of the dynamics to the timing of the control field. The information backflow occurs within specific time intervals determined by the environmental memory, and small variations in $\Omega(t)$ can shift the system away from these favorable windows. As a result, even the best pulse obtained across multiple runs produces only a weak and localized revival, consistent with the low accumulated non-Markovianity observed in Fig.~\ref{fig:total}.

\subsection{Optimized OCT pulses}

\begin{figure}[h]
\centering
\includegraphics[width=\linewidth]{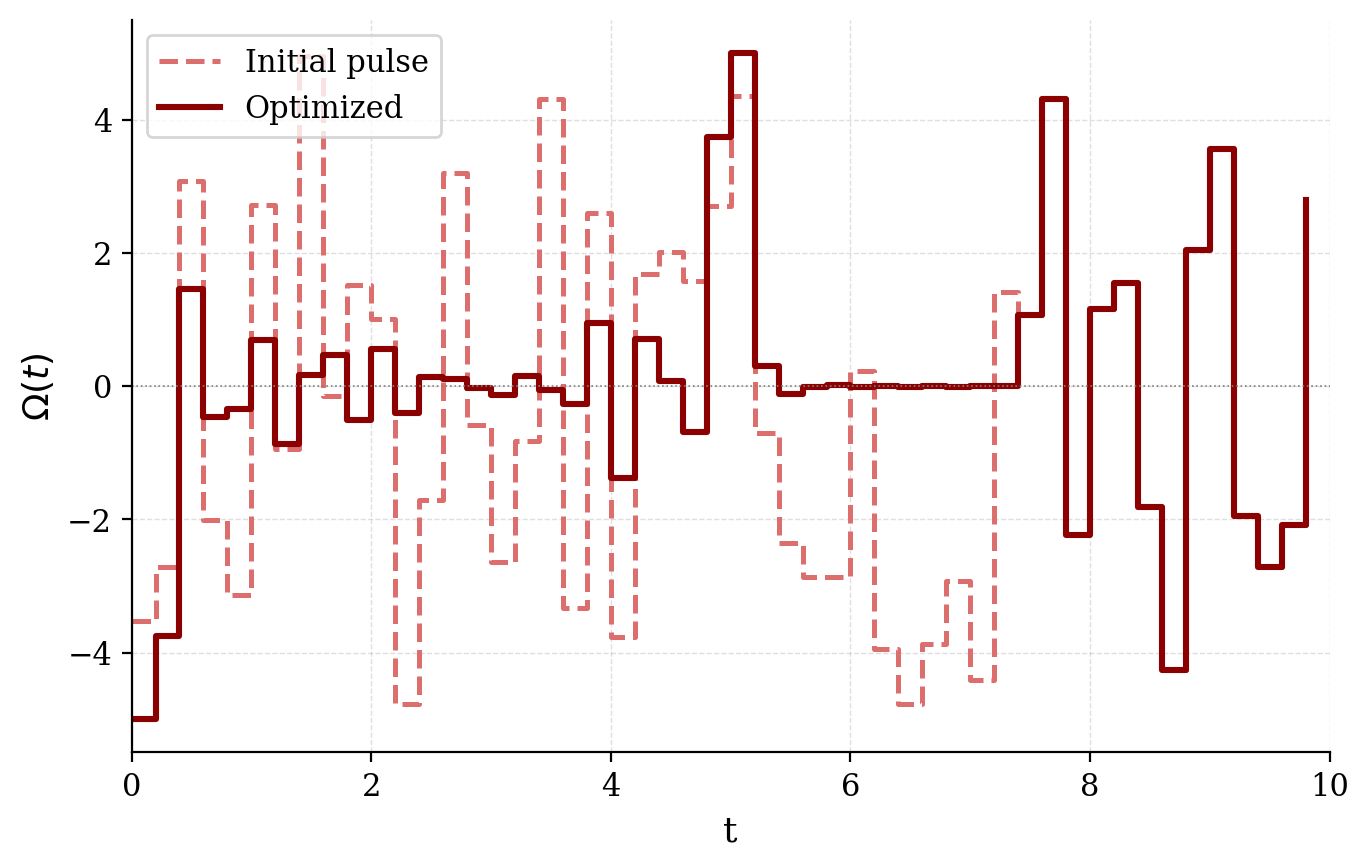}
\includegraphics[width=\linewidth]{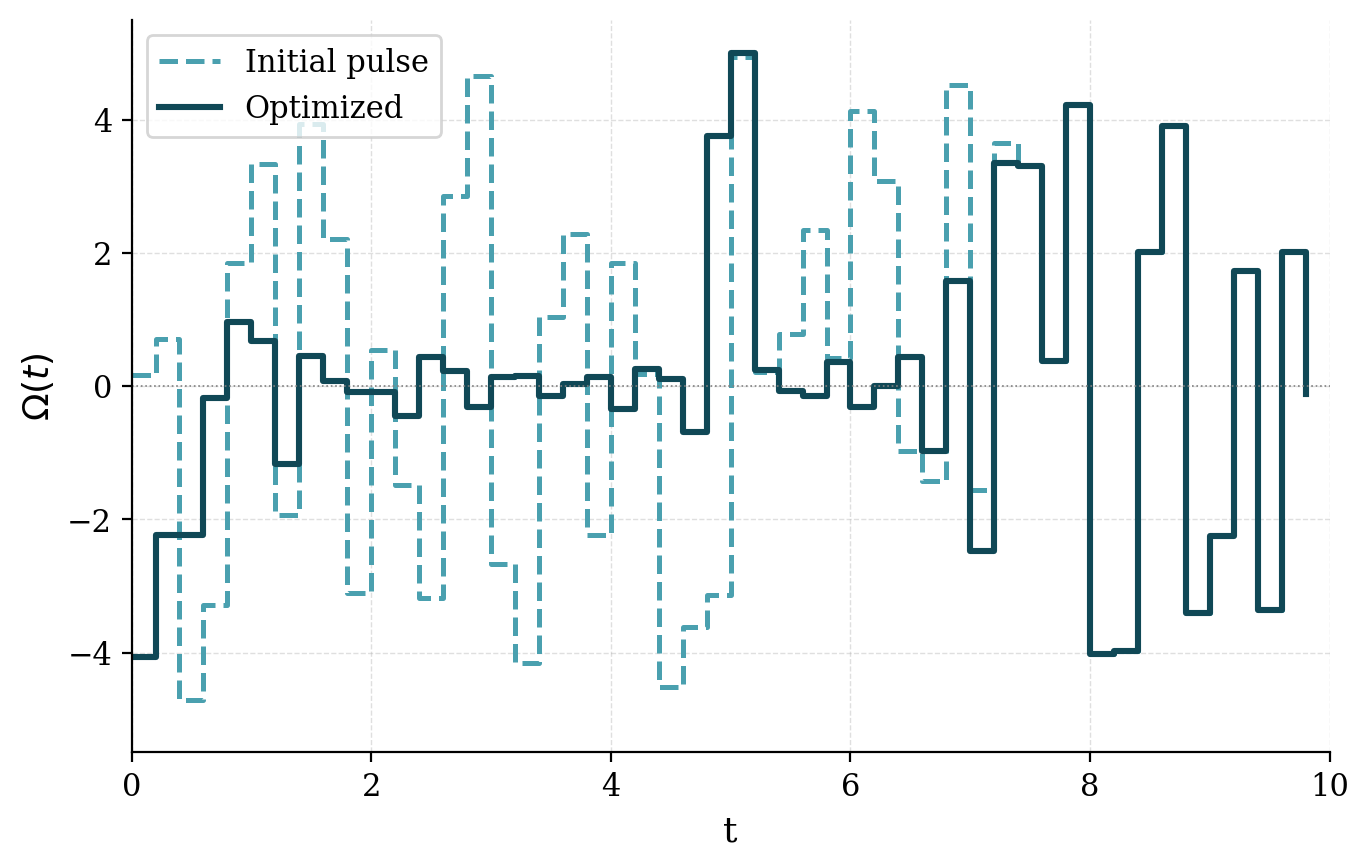}
\caption{Optimized control pulses $\Omega(t)$ obtained with Powell (top) and L-BFGS-B (bottom).}
\label{fig:oct_pulses}
\end{figure}

The optimized control pulses are shown in Fig.~\ref{fig:oct_pulses}.
Although both Powell and L-BFGS-B start from the same amplitude bounds and pulse discretization, they converge to slightly different pulse shapes.
This variability indicates that the OCT optimization does not lead to a unique control strategy, but rather to different solutions depending on the optimization path. However, despite these differences in the pulse profiles, the resulting enhancement of non-Markovianity remains limited, as evidenced by the low values of $\mathcal{N}_{\mathrm{Tot}}$ (Fig.~\ref{fig:total}) and the weak instantaneous revival (Fig.~\ref{fig:oct_local}).
This suggests that, in the presence of environmental memory, modifying the pulse shape alone is not sufficient to achieve strong information backflow within standard OCT approaches \cite{Fosel2018,Dalgaard2020}.

\section{Reinforcement-Learning Framework}

Reinforcement learning provides an alternative to OCT that bypasses the need for 
model gradients or adjoint backpropagation.  
Rather than optimizing a functional in a fixed control landscape, the agent 
learns directly from trajectory data generated during system evolution.  
At each step, it interacts with the quantum system, observes the resulting 
dynamics, receives rewards proportional to information backflow, and updates a 
policy $\pi(\Omega|s)$ that selects the control amplitude based on the observed 
state.  
This data-driven approach is naturally suited to non-Markovian environments, in 
which temporal correlations and memory revivals create rugged, highly 
non-convex landscapes that gradient-based methods often struggle to exploit 
consistently.

\subsection{RL environment formulation}

Before defining the observation and action spaces, we emphasize that the control 
field used here is \emph{stepwise constant in time}.  
During each interval $[t_k, t_{k+1})$ of duration $\Delta t$, the driving 
amplitude $\Omega_k$ is held fixed and updated only at discrete decision steps.  
The agent nonetheless operates in a \emph{continuous action space} 
($a_k \in \mathbb{R}$), which makes the task a continuous-action RL problem 
while retaining the standard piecewise-constant parametrization used in quantum 
control.

The observation provided to the agent at time step $(t_k)$ is defined as
\begin{equation*}
s_k = \big[ t_k/T, \rho_1(t_k), \rho_2(t_k), \Omega_{k-1} ,\big],
\end{equation*}
and contains only experimentally accessible quantities. It consists of the density matrices $(\rho_1(t_k))$ and $(\rho_2(t_k))$, evolved from two orthogonal initial states, the normalized time coordinate $(t_k/T)$, and the control amplitude applied at the previous time step $(\Omega_{k-1})$.
This representation ensures that the agent bases its decisions solely on observable system dynamics, without requiring any explicit knowledge of the environmental parameters (such as the decay rate $(\gamma(t)))$, whose effect is implicitly encoded in the evolution of the density matrices.
The action $(a_k \in \mathbb{R})$ corresponds to a continuous update of the control field, implemented as
\begin{equation}
\Omega_{k+1} = \mathrm{clip}\big(\Omega_k + a_k, \Omega_{\min}, \Omega_{\max}\big),
\end{equation}
where the clipping enforces the admissible amplitude bounds. This procedure generates a piecewise-constant control profile $(\Omega(t))$ over the full episode.


The reward function directly encodes the BLP information backflow:
\begin{equation}
r_k = \max(0, \dot{D}_k)
     \approx \max\!\left(0, \frac{D_{k+1}-D_k}{\Delta t}\right),
     \label{eq:10}
\end{equation}
encouraging the agent to detect, anticipate, and exploit intervals of positive 
$\dot{D}(t)$ without requiring any explicit knowledge of the analytic master 
equation.

\subsection{RL algorithms: SAC and PPO}

To learn optimal control strategies, we employ two complementary continuous-action RL algorithms:  
Soft Actor–Critic (SAC) and Proximal Policy Optimization (PPO).

SAC~\cite{Haarnoja2018} is an off-policy actor--critic method that has shown 
excellent performance in quantum control tasks~\cite{Niu2019,An2019}.  
It maximizes a stochastic objective with an entropy regularizer,
\begin{equation}
J(\pi)=\mathbb{E}_{(s,a)\sim\pi}\!\left[ Q(s,a)-\tau\log\pi(a|s)\right],
\label{eq:11}
\end{equation}
promoting exploratory yet purpose-driven behavior.  
Three features make SAC particularly powerful in non-Markovian settings:
\begin{enumerate}
    \item \emph{Maximum-entropy exploration} helps discover the coordinated 
          pulse patterns required to exploit intervals where $\gamma(t)$ becomes negative.
    \item \emph{Off-policy learning} enables extensive reuse of rare, highly 
          informative trajectories exhibiting strong memory revivals.
    \item \emph{Twin Q-networks} mitigate value overestimation in the 
          rugged, non-convex landscape generated by non-Markovian dynamics.
\end{enumerate}

To provide a robust and interpretable baseline, we also train a PPO agent.  
PPO is an on-policy actor--critic method whose updates are governed by the 
clipped surrogate objective
\begin{equation}
L^{\mathrm{CLIP}} =
\mathbb{E}\!\left[
\min\!\left(
r_k(\theta)\,\hat{A}_k,\;
\mathrm{clip}\!\big(r_k(\theta),\,1-\varepsilon,\,1+\varepsilon\big)\hat{A}_k
\right)
\right].
\label{eq:12}
\end{equation}

Here $r_k(\theta)=\pi_\theta(a_k|s_k)/\pi_{\theta_{\mathrm{old}}}(a_k|s_k)$ 
is the probability ratio comparing the new policy to the policy that generated 
the rollout, $\hat{A}_k$ is the advantage estimate indicating whether the action 
was better or worse than expected, and $\varepsilon$ is a clipping parameter 
that limits how far the policy is allowed to move in a single update.  
The clipping term prevents excessively large policy shifts, promoting stable and 
smooth control updates.  
PPO additionally employs generalized advantage estimation (GAE) to obtain 
low-variance advantage signals, which further encourages temporally coherent 
adjustments to the pulse amplitude.

Although PPO is generally less sample-efficient than SAC and explores more conservatively, it is well known for stable and reliable convergence in continuous-control tasks. Its incremental update mechanism and implicit regularization tend to produce smoother policy variations over time, which can be advantageous when the control landscape exhibits structured temporal features.

In contrast, SAC adopts an off-policy learning strategy with entropy regularization, promoting broader exploration of the action space. This typically leads to more flexible and rapidly varying control policies. Comparing PPO and SAC therefore provides insight into how different learning mechanisms—on-policy versus off-policy—interact with the temporal complexity of memory effects in non-Markovian quantum systems.

\subsection{RL workflow}

\begin{figure}[h]
    \centering
    \includegraphics[width=\linewidth]{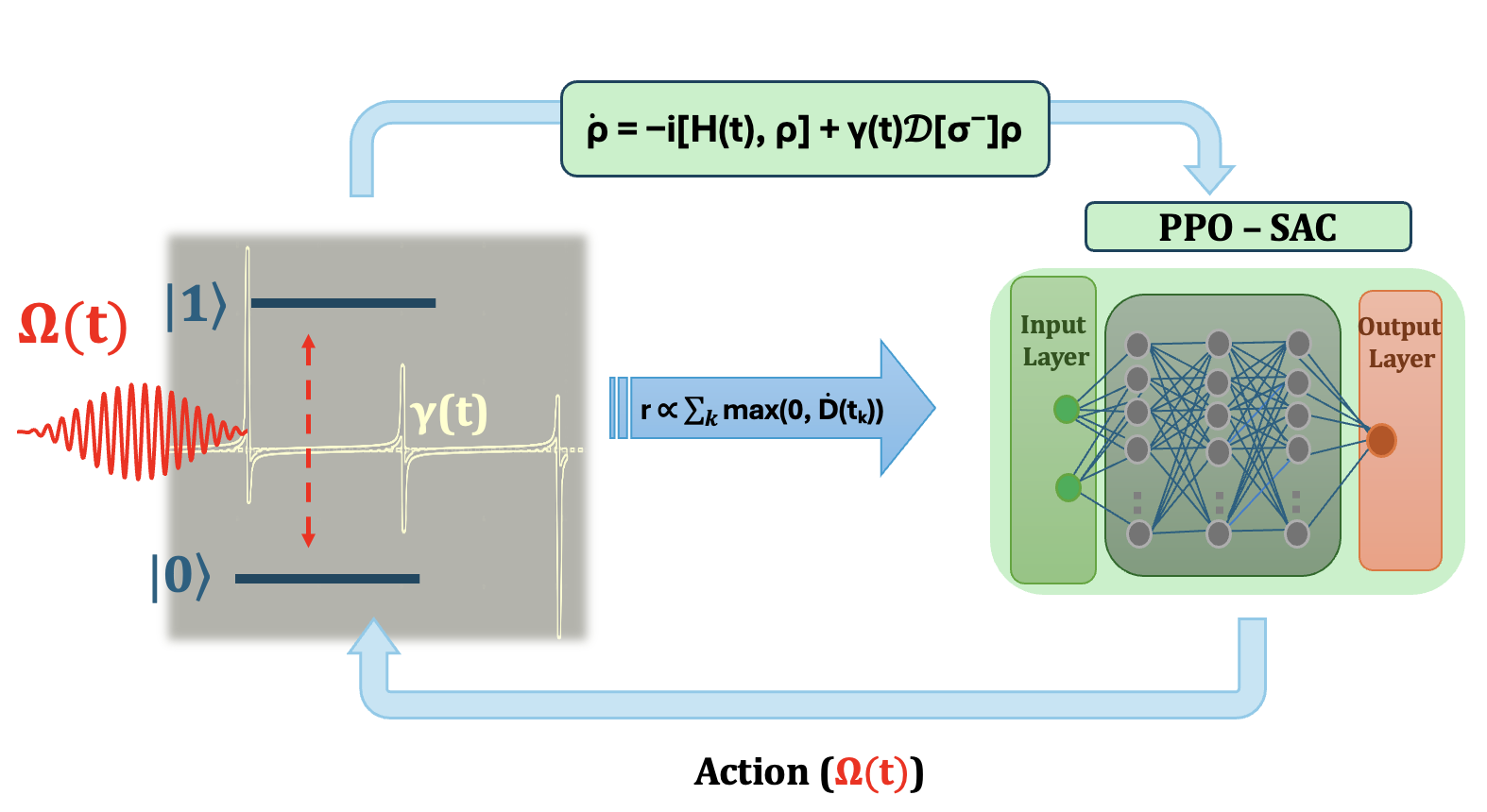}
    \caption{RL control loop for maximizing
    non-Markovianity.}
    \label{fig:RLloop}
\end{figure}

A schematic overview of the RL loop is shown in Fig.~\ref{fig:RLloop}.  
Both SAC and PPO interact with the environment through the following sequence:

\begin{enumerate}
    \item Initialize $\rho_1(0)$, $\rho_2(0)$, and a random initial amplitude $\Omega_0$.
    \item Apply the action $a_k$ and propagate the master equation for one step $\Delta t$.
    \item Compute $D_k$, $\dot D_k$, and the reward $r_k$.
    \item Store the transition $(s_k, a_k, r_k, s_{k+1})$:  
          SAC uses a replay buffer, while PPO uses on-policy rollouts.
    \item Update the actor and critics using gradient descent according to the chosen RL algorithm.
\end{enumerate}

\section{Results and Discussion}
\subsection{OCT versus RL: peaks versus area}

\begin{figure}[h]
\centering
\includegraphics[width=\linewidth]{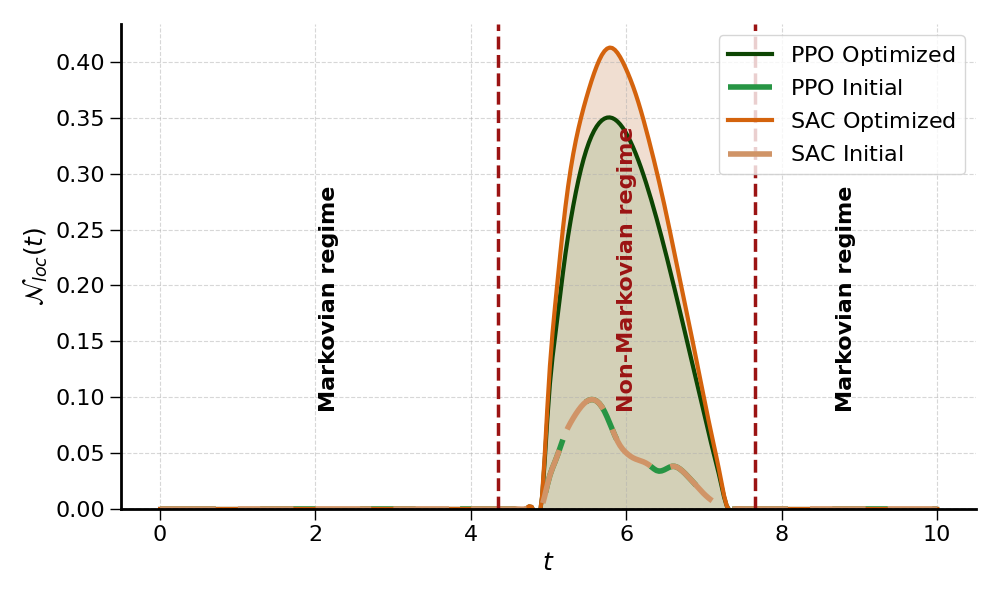}
\caption{Instantaneous non-Markovianity $\mathcal{N}_{\mathrm{loc}}(t)$ corresponding to the best pulse found over 5 independent runs, for SAC (orange) and PPO (green). Solid lines denote optimized pulses, while dashed lines denote the initial random pulses. Shaded regions indicate the non-Markovian regime where $\mathcal{N}_{\mathrm{loc}}(t) > 0$.}
\label{fig:bars_RL}
\end{figure}

Figure~\ref{fig:oct_local} summarizes the behavior of the Powell and L-BFGS-B OCT algorithms.
Standard OCT increases both the height and the width of the main local non-Markovianity peak
(solid lines) compared with the uncontrolled dynamics (dashed line).
However, this enhancement remains confined to a relatively narrow time window
associated with the first memory interval.
In particular, the dominant OCT-induced revival occurs around $t \simeq 4-8$
and reaches a maximal instantaneous non-Markovianity
$\mathcal{N}_{\mathrm{loc}}^{\max} \approx 0.0065$, consistent with the modest plateau
value $\mathcal{N}_{\mathrm{Tot}} \simeq 0.009$ observed in Fig.~\ref{fig:total}.
In contrast, the RL-controlled dynamics shown in
Fig.~\ref{fig:bars_RL} exhibit a qualitatively different behavior.
Both PPO and SAC generate substantially larger instantaneous revivals within the same general time region,
with SAC reaching peak values
$\mathcal{N}_{\mathrm{loc}}^{\max} \approx 0.42$, far exceeding those obtained with OCT,
while PPO achieves $\mathcal{N}_{\mathrm{loc}}^{\max} \approx 0.3$.
This demonstrates that RL is able to significantly amplify the information backflow beyond what is achievable with standard OCT.
At first sight, although standard OCT enhances the main backflow event, its effect remains weak.
RL, by contrast, produces a much stronger instantaneous backflow,
which directly translates into a substantially higher integrated non-Markovianity.
This distinction is central to interpreting our results.
Standard OCT produces a weak and localized burst of information backflow,
whereas RL achieves a much stronger amplification of the backflow dynamics.
Although the decay rate $\gamma(t)$ becomes negative again at later times,
no pronounced revival is observed around $t \simeq 9-10$, as the distinguishability
between the trajectories has already been significantly reduced.
A negative decay rate is therefore a necessary but not sufficient condition for
information backflow.

For clarity, we emphasize that both standard OCT and RL are formulated to optimize the same physical objective, namely the total non-Markovianity quantified by the BLP measure. In our implementation, the standard OCT cost function corresponds to the discrete-time sum of positive increments of the trace distance, which is identical to the cumulative reward used in RL.
Therefore, the comparison between standard OCT and RL does not rely on different optimization targets, but rather on the distinct ways in which these methods explore and exploit the control landscape. While standard OCT performs a global optimization of the full functional, RL operates through sequential, trajectory-based updates that explicitly reinforce positive increments of the trace distance at each step.
This difference in optimization strategy leads to qualitatively distinct control behaviors, particularly in non-Markovian regimes where the temporal structure of memory effects plays a central role.

\subsection{Global performance comparison}

\begin{figure}[h]
    \centering
    \includegraphics[width=\linewidth]{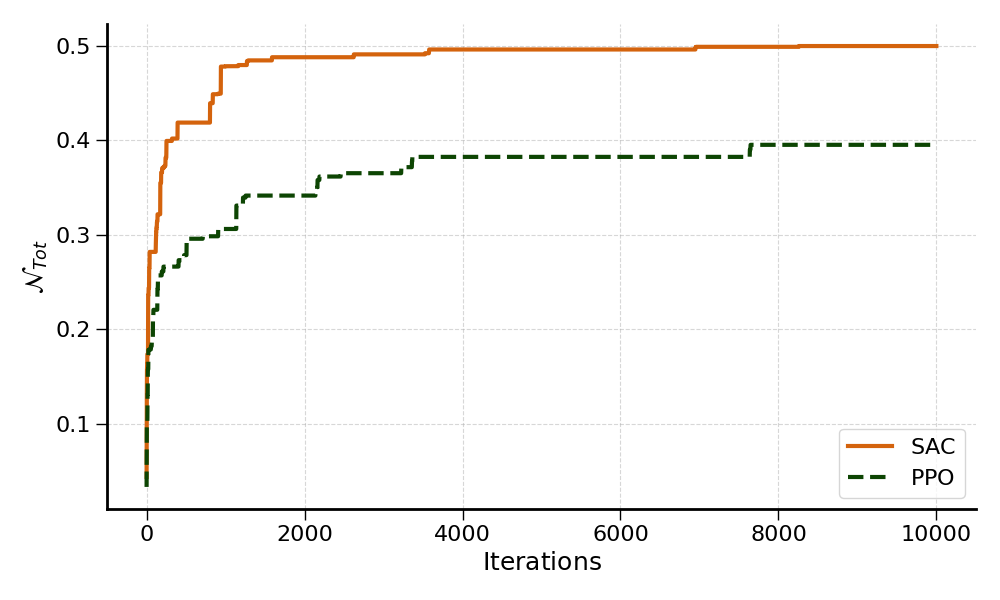}
\caption{
Convergence history of the total non-Markovianity
$\mathcal{N}_{\mathrm{Tot}}$ averaged over 5 independent runs. SAC (solid orange) and PPO (dashed green).
Both agents rapidly improve their performance, with SAC reaching the highest final value.}
    \label{fig:rl_local}
\end{figure}

The global comparison is shown in Fig.~\ref{fig:rl_local}, where we track the evolution of the total non-Markovianity $\mathcal{N}_{\mathrm{Tot}}$ during training for PPO and SAC. Both agents rapidly escape the low-performance regime associated with random initial pulses: within the first few hundred iterations, $\mathcal{N}_{\mathrm{Tot}}$ already exceeds $0.2$ for both methods.
After this initial rise, their learning dynamics diverge. PPO exhibits a staircase-like progression characterized by abrupt policy improvements, reaching a plateau around $\mathcal{N}_{\mathrm{Tot}} \simeq 0.4$ after approximately $8\times10^3$ iterations. In contrast, SAC stabilizes early and continues to improve through smaller but more frequent updates, ultimately reaching higher values $\mathcal{N}_{\mathrm{Tot}} \simeq 0.48 - 0.5$.

The resulting hierarchy is:
\begin{enumerate}
\item SAC attains the largest total non-Markovianity
$\mathcal{N}_{\mathrm{Tot}} \simeq 0.5$.
\item PPO reaches a slightly lower but still strongly enhanced value
$\mathcal{N}_{\mathrm{Tot}} \simeq 0.4$.
\item OCT improves upon the uncontrolled dynamics but remains far below the
RL policies in total $\mathcal{N}_{\mathrm{Tot}}$.
\end{enumerate}

At first sight, one might expect that amplifying the dominant backflow event, as standard OCT does, would be sufficient to maximize the total non-Markovianity. However, Fig.~\ref{fig:rl_local} shows that this intuition is incomplete.
While standard OCT produces only a weak and localized backflow, RL achieves a much stronger amplification of this dominant event, leading to a substantially larger integrated non-Markovianity. This difference arises from the distinct optimization mechanisms. Again standard OCT relies on global gradient-based updates and becomes trapped in low-performing regions of the control landscape, whereas RL operates through sequential, trajectory-based updates that reinforce positive increments of the trace distance.
As a result, PPO and SAC progressively discover control strategies that enhance information backflow more effectively, with SAC benefiting from broader exploration and achieving the highest performance.

\subsection{RL pulse shaping}
\begin{figure}[H]
\centering
\includegraphics[width=\linewidth]{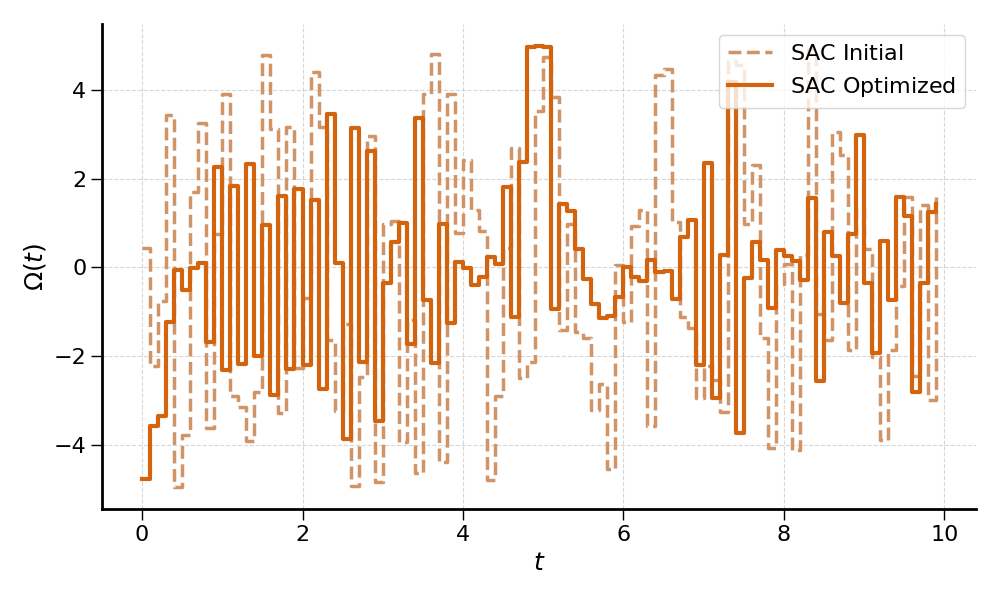}
\includegraphics[width=\linewidth]{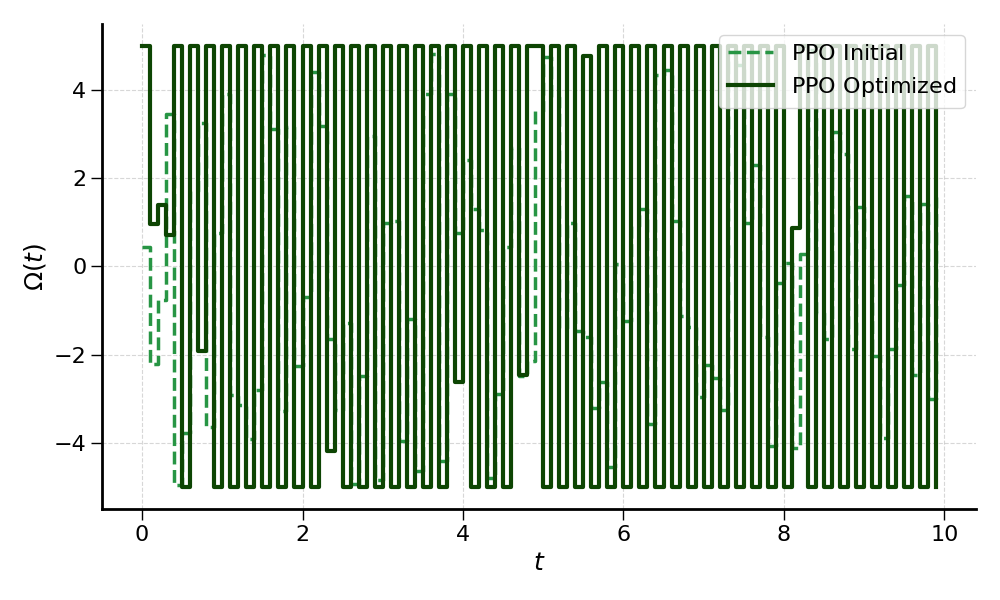}
\caption{ Optimized control pulses corresponding to the best-performing runs obtained with SAC (top) and PPO (bottom), compared to their respective initial random pulses (dashed).}
\label{fig:rl_pulses}
\end{figure}

To understand how the different RL algorithms shape the control field, we compare in Fig.~\ref{fig:rl_pulses} the optimized pulses corresponding to the best-performing runs among the independent training realizations for SAC and PPO. Although both start from similar random initial pulses, their learned solutions exhibit systematic differences once training converges.
The SAC-optimized pulse (top panel) displays pronounced small-scale fluctuations and frequent variations in amplitude. This behavior reflects the entropy-regularized objective of SAC, which promotes broader exploration of the control landscape and leads to highly modulated control profiles.
By contrast, the PPO pulse (bottom panel) is noticeably smoother and more regular. Its amplitude evolves through longer and more coherent segments, with fewer rapid variations between successive time steps. This behavior is consistent with PPO’s clipped-surrogate update rule, which limits large policy changes and favors more stable control strategies.
These differences in pulse structure are reflected in the resulting dynamics. While both approaches significantly enhance the information backflow compared to standard OCT, SAC achieves a larger instantaneous non-Markovianity and a higher total $\mathcal{N}_{\mathrm{Tot}}$, as shown in Fig.~\ref{fig:rl_local}.
This indicates that the strongly modulated SAC pulses, despite their apparent irregularity, are more effective at enhancing backflow dynamics. In contrast, the smoother PPO pulses lead to a robust but comparatively less optimal exploitation of the non-Markovian behavior.
Overall, this comparison highlights how different learning strategies translate into distinct control profiles and performance levels, with SAC benefiting from a broader exploration of the control landscape and achieving the highest non-Markovianity.

\subsection{Statistical Robustness}
\begin{figure}[H]
\centering
\includegraphics[width=\linewidth]{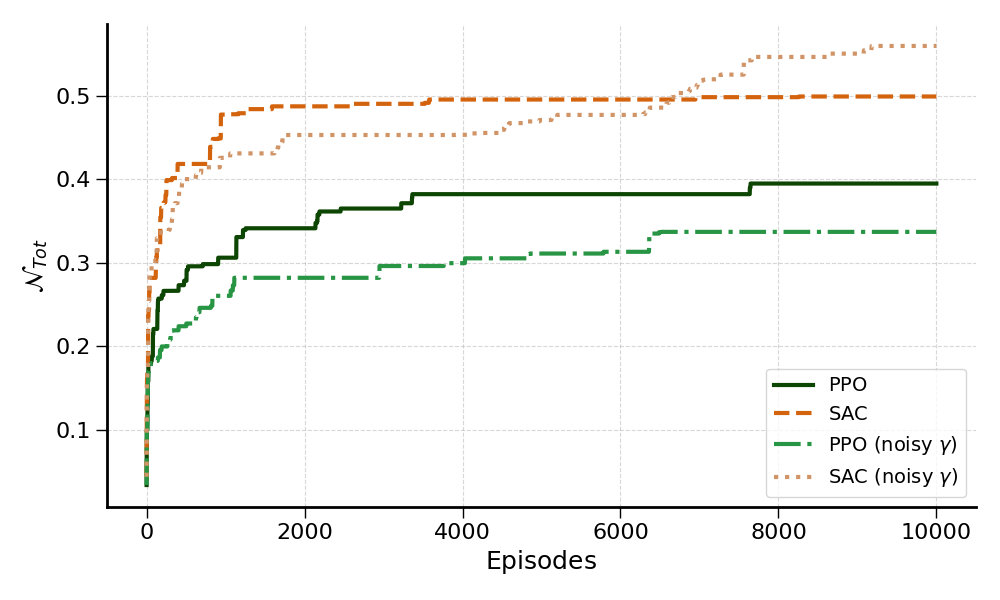}
\caption{Evolution of the total non-Markovianity $\mathcal{N}_{\mathrm{Tot}}$ as a function of training iterations, averaged over 5 independent runs with different random seeds, for the ideal and noisy $\gamma(t)$ cases.}
\label{mean_ntot}
\end{figure}

To assess the statistical robustness of the reinforcement learning agents and account for the variability inherent to stochastic optimization, each algorithm (PPO and SAC) was trained over 5 independent runs, each initialized with a distinct random seed. Figure~\ref{mean_ntot} shows the evolution of the total non-Markovianity $\mathcal{N}_{\mathrm{Tot}}$ as a function of training iterations, averaged over all runs.
As shown in Fig.~\ref{mean_ntot}, both PPO and SAC rapidly escape the low-performance regime associated with random initial pulses. Within the first few hundred iterations, both methods reach $\mathcal{N}_{\mathrm{Tot}} \gtrsim 0.2$. Their learning dynamics then diverge. SAC quickly stabilizes around $\mathcal{N}_{\mathrm{Tot}} \simeq 0.45$ and continues to improve gradually, reaching a final plateau near $\mathcal{N}_{\mathrm{Tot}} \simeq 0.5$. In contrast, PPO converges more slowly and saturates around $\mathcal{N}_{\mathrm{Tot}} \simeq 0.4$, remaining consistently below SAC throughout training.
The smoothness of the averaged curves indicates that run-to-run fluctuations are small compared to the separation between the methods. At convergence, the performance gap between SAC and PPO is of the order $\Delta \mathcal{N}_{\mathrm{Tot}} \sim 0.1$, confirming that the superiority of SAC is statistically significant and not due to a particular realization. The initial control pulse in each run is drawn uniformly at random from $\Omega(t) \in [-5,5]$, ensuring unbiased initialization.
We further compare the ideal case (without fluctuations in $\gamma(t)$) and the noisy case (with stochastic perturbations), using identical statistical conditions. The noise is modeled as:
$\gamma_{\mathrm{noisy}}(t) = \gamma(t) + \epsilon(t), \qquad \epsilon(t) \sim \mathcal{N}(0,\sigma^2),$
with $\sigma = 10^{-4}$.
As seen in Fig.~\ref{mean_ntot}, the effect of noise differs for the two algorithms. For PPO, the presence of noise leads to a slight reduction in the final performance, with $\mathcal{N}_{\mathrm{Tot}}$ decreasing by approximately $0.05$. In contrast, SAC maintains its performance and even exhibits a small increase in the final value of $\mathcal{N}_{\mathrm{Tot}}$. Importantly, the overall convergence behavior and the relative ranking of the methods remain unchanged.
This asymmetric response reflects the different learning dynamics of the two algorithms. PPO, which converges toward smoother and more deterministic policies, is more sensitive to perturbations in the system dynamics. By contrast, SAC incorporates stochastic exploration through entropy regularization, enabling it to explore a broader region of the control landscape and making it more resilient to fluctuations. As a result, SAC not only achieves higher performance in the ideal setting but also exhibits stronger robustness to environmental noise.

\subsection{Hyperparameter Sensitivity Analysis}

To assess the robustness of the proposed optimization framework, we conduct a systematic sensitivity analysis with respect to three key hyperparameters: the number of time steps $N_{\mathrm{steps}}$, the learning rate $\alpha$, and the discount factor $\eta$. Each parameter is varied independently while keeping the others fixed

\begin{figure}[H]
\centering
\includegraphics[width=\linewidth]{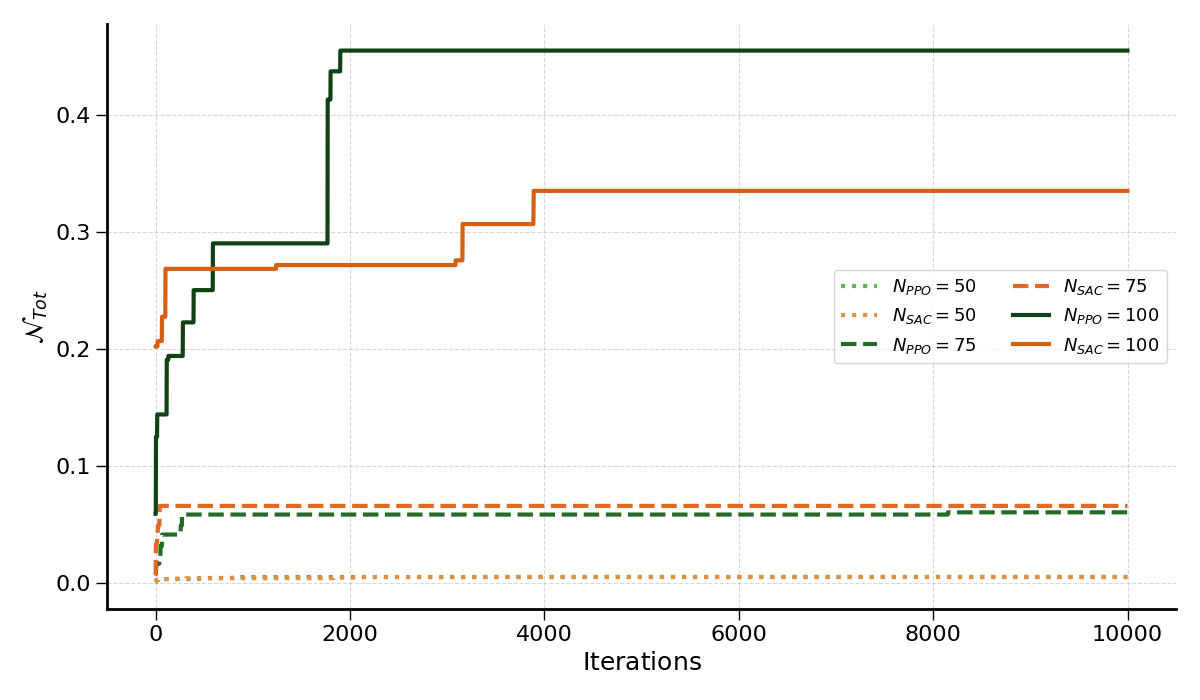}
\caption{ Effect of the temporal discretization $N_{\mathrm{steps}}$ on the training convergence of
$\mathcal{N}_{\mathrm{Tot}}$ for SAC and PPO. Increasing $N_{\mathrm{steps}}$ improves the achievable performance but slows down convergence due to the higher dimensionality of the control space.}
\label{fig:steps}
\end{figure}

\paragraph{Number of time steps.}
The discretization of the control pulse into $N_{\mathrm{steps}}$ time steps determines both the temporal resolution of the pulse and the episode length in the RL framework. We evaluate $N_{\mathrm{steps}} \in {50, 75, 150}$ and report the resulting convergence behavior in Fig.~\ref{fig:steps}.
As shown in Fig.~\ref{fig:steps}, increasing $N_{\mathrm{steps}}$ improves the achievable performance, with both PPO and SAC reaching higher values of $\mathcal{N}_{\mathrm{Tot}}$ for finer discretizations. This reflects the ability of the agents to resolve sharper temporal features in the control pulse. However, this improvement comes at the cost of slower convergence due to the increased dimensionality of the control problem. The choice $N_{\mathrm{steps}} = 100$ provides a good compromise between performance and computational efficiency.

\begin{figure}[H]
\centering
\includegraphics[width=\linewidth]{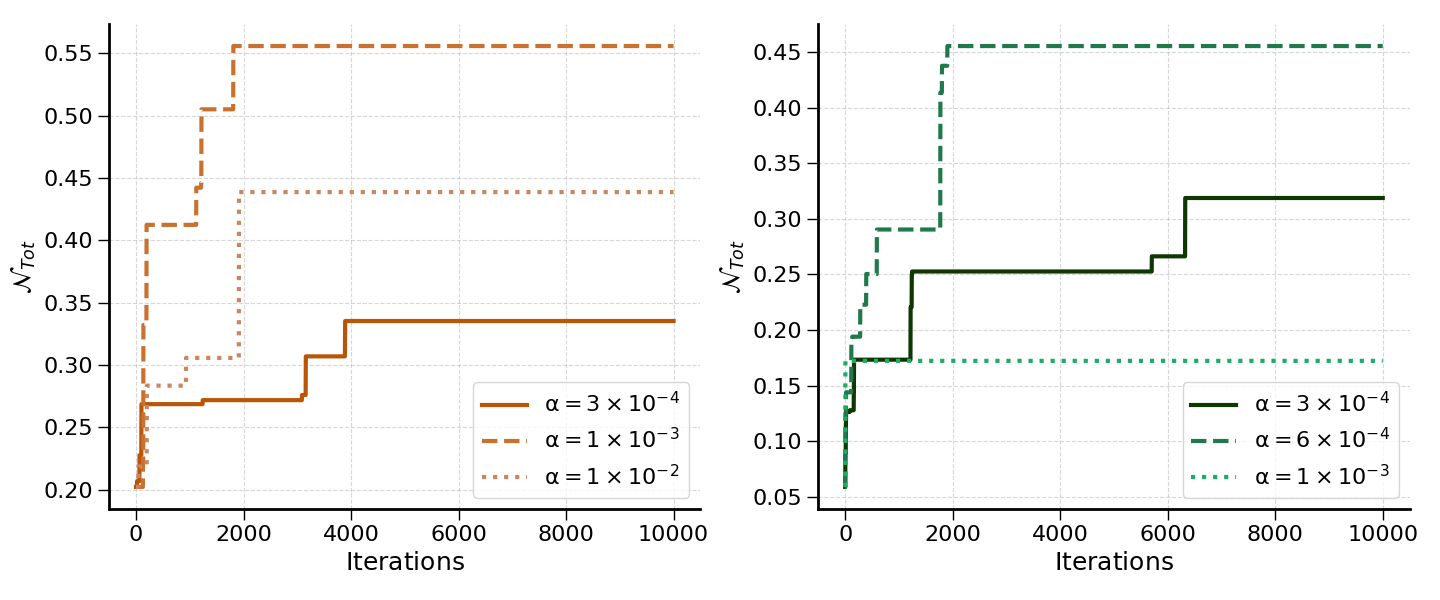}
\caption{ Effect of the learning rate $\alpha$ on the training convergence of $\mathcal{N}_{\mathrm{Tot}}$ for PPO (left) and SAC (right). Intermediate values provide the best balance between convergence speed and stability.}
\label{lr}
\end{figure}

\paragraph{Learning rate.}
The learning rate $\alpha$ controls the step size of the policy update and therefore strongly influences training stability. We explore $\alpha \in {10^{-4}, 3\times10^{-4}, 6\times10^{-4}}$ for PPO and $\alpha \in {10^{-2}, 10^{-3}, 3\times10^{-4}}$ for SAC, as shown in Fig.~\ref{lr}.
Figure~\ref{lr} shows that both algorithms exhibit a clear optimal range of learning rates. For PPO, intermediate values ($\alpha \approx 3\times10^{-4}$) yield the best convergence, while too large values lead to instability and too small values slow down learning. For SAC, larger learning rates allow faster initial improvement, but overly large values can introduce fluctuations in the training dynamics. Importantly, across all tested values, SAC consistently reaches higher final $\mathcal{N}_{\mathrm{Tot}}$ than PPO, indicating that the observed performance hierarchy is robust with respect to this hyperparameter.

\begin{figure}[H]
\centering
\includegraphics[width=\linewidth]{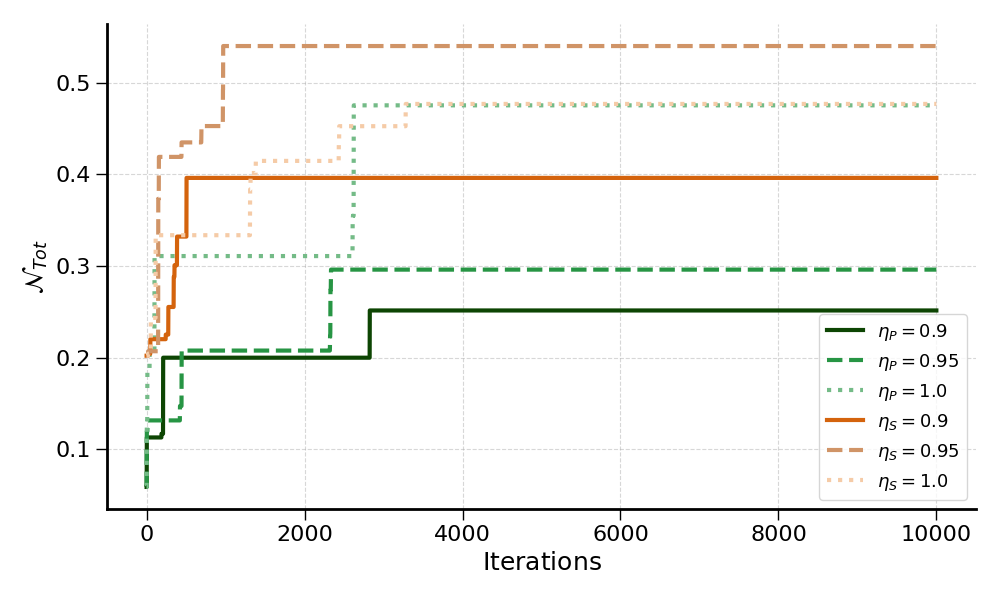}
\caption{ Training convergence of $\mathcal{N}_{\mathrm{Tot}}$ for PPO and SAC under different discount factors $\eta \in {0.90, 0.95, 1.0}$. Values close to unity yield the best performance, consistent with the cumulative nature of the objective.}
\label{fig:discount}
\end{figure}

\paragraph{Discount factor.}
The discount factor $\eta \in [0,1]$ controls the balance between immediate and long-term rewards. Since the total non-Markovianity $\mathcal{N}_{\mathrm{Tot}}$ is an accumulated quantity over the full trajectory, the choice $\eta = 1$ is physically well motivated. Nevertheless, we investigate $\eta \in {0.90, 0.95, 1.0}$ to quantify the effect of discounting, as shown in Fig.~\ref{fig:discount}.
As illustrated in Fig.~\ref{fig:discount}, decreasing $\eta$ leads to slower convergence and lower final values of $\mathcal{N}_{\mathrm{Tot}}$ for both PPO and SAC. This behavior is expected, as discounting reduces the contribution of later-time backflow to the objective function. In contrast, values close to unity allow the agents to fully exploit the accumulated nature of non-Markovianity, resulting in optimal performance. Again, SAC maintains a consistent advantage over PPO across all tested values of $\eta$.

Overall, this analysis demonstrates that the performance of both algorithms is robust over a broad range of hyperparameters, and that the superiority of SAC is not the result of a particular parameter choice.

\subsection{Comparative Roles and Trade-offs of
PPO, SAC, and standard OCT}

Drawing on both the learning curves and the optimized pulse structures, we can now clarify the distinct roles, strengths, and limitations of standard OCT, PPO, and SAC in controlling non-Markovian dynamics.

Standard OCT enhances the dominant backflow event but only weakly. Its gradient-based updates reinforce the most prominent local feature of the landscape, leading to a small and localized peak in $\mathcal{N}_{\mathrm{loc}}(t)$. As a result, standard OCT quickly saturates and remains limited in its ability to increase the total non-Markovianity $\mathcal{N}_{\mathrm{Tot}}$. This behavior reflects the difficulty of optimizing a history-dependent objective within a complex control landscape shaped by environmental memory.

RL uncovers a qualitatively different strategy. By receiving reward whenever $\dot D(t)>0$, PPO and SAC progressively learn control policies that enhance the dominant backflow event more effectively. Rather than simply amplifying this event marginally, RL significantly increases its amplitude, leading to much larger instantaneous non-Markovianity and, consequently, a substantially higher integrated $\mathcal{N}_{\mathrm{Tot}}$.

Within the RL family, PPO and SAC exhibit distinct behaviors that reflect their algorithmic design. PPO, based on on-policy updates with clipped objectives, tends to generate smoother and more stable control pulses. This leads to robust and consistent performance, with $\mathcal{N}_{\mathrm{Tot}} \simeq 0.4$.
SAC, by contrast, relies on an off-policy learning scheme with entropy regularization, promoting broader exploration of the control landscape. This results in more strongly modulated control pulses, which, despite their apparent irregularity, prove more effective at enhancing the backflow dynamics. As a result, SAC achieves the highest performance, reaching $\mathcal{N}_{\mathrm{Tot}} \simeq 0.5$.

In summary, standard OCT produces only a weak and localized enhancement of the backflow, whereas reinforcement learning significantly amplifies this dominant event. Among RL methods, PPO provides stable and smooth control strategies, while SAC leverages enhanced exploration to achieve the highest non-Markovianity. This demonstrates that the advantage of RL and in particular SAC stems from its ability to more effectively exploit the dynamical structure of non-Markovian systems.

\section{Conclusion}
We have developed a RL framework for enhancing non-Markovianity in a driven open quantum system by maximizing the BLP measure through direct trajectory-based feedback. Unlike gradient-based standard OCT, which is limited by the complexity of the control landscape and yields only a weak amplification of the dominant backflow event, RL identifies control strategies that significantly enhance the information backflow dynamics.
Our numerical results establish a clear hierarchy among the methods. Both PPO and SAC achieve substantially larger total non-Markovianity than Powell and L-BFGS-B standard OCT. In the parameter regime considered here, RL outperforms standard OCT not only in terms of the integrated measure $\mathcal{N}_{\mathrm{Tot}}$, but also in the amplitude of the instantaneous backflow. In particular, SAC consistently achieves the highest performance, reaching $\mathcal{N}_{\mathrm{Tot}} \simeq 0.5$, while PPO attains slightly lower but still strongly enhanced values around $\mathcal{N}_{\mathrm{Tot}} \simeq 0.4$.
The comparison between PPO and SAC highlights the impact of different learning strategies. PPO, based on stable on-policy updates, generates smoother and more regular control pulses, leading to robust but moderately optimal performance. SAC, by contrast, leverages entropy-regularized off-policy learning to explore a broader region of the control landscape. This results in more strongly modulated control pulses, which prove more effective at enhancing the dominant backflow event and yield the highest non-Markovianity.
These findings highlight a key advantage of RL: rather than marginally improving the existing backflow dynamics, RL is able to significantly amplify the dominant information backflow event, leading to a much larger integrated non-Markovianity. This demonstrates that model-free RL controllers can effectively exploit the dynamical structure of non-Markovian systems beyond the capabilities of standard gradient-based OCT.
Finally, the ability of RL agents to learn control strategies directly from trajectory-based feedback, without requiring explicit gradients or detailed modeling of the environment, makes them promising candidates for experimental implementation. Future directions include extending this approach to multi-qubit systems, combining RL with bath engineering techniques, and integrating real-time feedback protocols to actively control memory effects in quantum technologies.

\section*{Acknowledgments}

S.G. acknowledges the financial support of the National 
Center for Scientific and Technical Research (CNRST)
through the “PhD-Associate Scholarship–PASS” program.

\appendix
\section{Computational Models and Integration Schemes}

We summarize here the different OCT and RL routines used in this work. All algorithms operate on the vector of control amplitudes
$\boldsymbol{\Omega}=(\Omega_1,\dots,\Omega_{N_c})$ defining the piecewise
constant pulse $\Omega(t)$, and attempt to maximize the total
non-Markovianity $\mathcal{N}_{Tot}(\boldsymbol{\Omega})$ obtained from a full
master-equation propagation.

\begin{figure*}[h]
\centering
\fbox{%
\begin{minipage}{0.95\textwidth}
\textbf{Algorithm 1: Powell OCT}
\medskip

\textbf{Input:} initial control vector
$\boldsymbol{\Omega}^{(0)}\in[\Omega_{\min},\Omega_{\max}]^{N_c}$,
initial set of search directions
$\{\mathbf{d}_1,\dots,\mathbf{d}_{N_c}\}$ (e.g. canonical basis),
maximum number of outer iterations $N_{\rm iter}$.

\medskip
\textbf{Objective:} maximize
$\mathcal{N}(\boldsymbol{\Omega})=\sum_k \max(0,\dot D(t_k))\Delta t$.
\medskip

\begin{enumerate}
\item \textbf{Initialize} iteration counter $m=0$ and evaluate
$\mathcal{N}^{(0)} = \mathcal{N}(\boldsymbol{\Omega}^{(0)})$ by:
  \begin{enumerate}
  \item constructing the piecewise-constant pulse $\Omega(t)$ from
  $\boldsymbol{\Omega}^{(0)}$;
  \item propagating the master equation over $[0,T]$;
  \item computing $\dot D(t_k)$, $\mathcal{N}_{\rm loc}(t_k)$ and
  $\mathcal{N}^{(0)}$.
  \end{enumerate}

\item \textbf{Repeat} for $m=0,1,2,\dots,N_{\rm iter}-1$:
  \begin{enumerate}
  \item Set $\boldsymbol{\Omega}_{\rm start}=\boldsymbol{\Omega}^{(m)}$
  and $\mathcal{N}_{\rm start}=\mathcal{N}^{(m)}$.

  \item \textbf{Cyclic line searches:}
    \begin{enumerate}
    \item For $j=1,\dots,N_c$:
      \begin{enumerate}
      \item Define a 1D search along direction $\mathbf{d}_j$:
      \[
        \boldsymbol{\Omega}(\lambda)
        = \boldsymbol{\Omega}^{(m)} + \lambda\,\mathbf{d}_j.
      \]
      \item Perform a line search over $\lambda$ (respecting the amplitude
      bounds) to find
      \[
      \lambda^\star = \arg\max_{\lambda}\,
      \mathcal{N}(\boldsymbol{\Omega}(\lambda)),
      \]
      where each evaluation of $\mathcal{N}$ requires a full master-equation
      propagation.
      \item Update the control:
      \[
      \boldsymbol{\Omega}^{(m)} \leftarrow \boldsymbol{\Omega}(\lambda^\star),
      \quad
      \mathcal{N}^{(m)} \leftarrow \mathcal{N}(\boldsymbol{\Omega}^{(m)}).
      \]
      \end{enumerate}
    \end{enumerate}

  \item \textbf{Direction update:}
    \begin{itemize}
      \item Compute the net displacement
      $\mathbf{d}_{\rm new}=\boldsymbol{\Omega}^{(m)}-\boldsymbol{\Omega}_{\rm start}$.
      \item Optionally discard the oldest direction and append $\mathbf{d}_{\rm new}$
      to $\{\mathbf{d}_1,\dots,\mathbf{d}_{N_c}\}$, following Powell’s rule,
      to capture curvature information.
    \end{itemize}

  \item \textbf{Stopping criterion:}
    \begin{itemize}
      \item If $\mathcal{N}^{(m)}-\mathcal{N}_{\rm start}$ or
      $\|\mathbf{d}_{\rm new}\|$ falls below a small threshold, stop.
    \end{itemize}
  \end{enumerate}
\end{enumerate}

\medskip
\textbf{Output:} optimized control vector $\boldsymbol{\Omega}^\star$ and
pulse $\Omega^\star(t)$ corresponding to a (possibly local) maximum of
$\mathcal{N}$.
\end{minipage}
}
\caption{Pseudocode for Powell's derivative-free optimal-control algorithm
applied to the maximization of total non-Markovianity.}
\label{alg:powell}
\end{figure*}

\begin{figure*}[h]
\centering
\fbox{%
\begin{minipage}{0.95\textwidth}
\textbf{Algorithm 2: L-BFGS-B OCT}
\medskip

\textbf{Input:} initial control vector
$\boldsymbol{\Omega}^{(0)}\in[\Omega_{\min},\Omega_{\max}]^{N_c}$,
step-size and convergence tolerances, finite-difference step $\epsilon$,
maximum number of iterations $N_{\rm iter}$.

\medskip
\textbf{Objective:} maximize
$\mathcal{N}(\boldsymbol{\Omega})$ under bound constraints
$\Omega_{\min}\le\Omega_j\le\Omega_{\max}$.
\medskip

\begin{enumerate}
\item \textbf{Initialize} iteration $m=0$ and evaluate
$\mathcal{N}^{(0)}=\mathcal{N}(\boldsymbol{\Omega}^{(0)})$ via a full
master-equation propagation.

\item \textbf{Repeat} for $m=0,1,2,\dots,N_{\rm iter}-1$:
  \begin{enumerate}
  \item \textbf{Gradient estimation (finite differences):}
    \begin{enumerate}
    \item For $j=1,\dots,N_c$:
      \begin{enumerate}
      \item Construct a perturbed control vector
      $\boldsymbol{\Omega}^{(m)}_{(j)}$ by replacing
      $\Omega_j^{(m)}\to\Omega_j^{(m)}+\epsilon$ (and clipping to bounds).
      \item Evaluate the perturbed cost
      $\mathcal{N}_{(j)}=\mathcal{N}(\boldsymbol{\Omega}^{(m)}_{(j)})$ by
      a full propagation.
      \item Form the numerical gradient component
      \[
      g_j^{(m)} =
      \frac{\mathcal{N}_{(j)}-\mathcal{N}^{(m)}}{\epsilon}.
      \]
      \end{enumerate}
    \item Assemble the gradient vector
    $\mathbf{g}^{(m)}=(g_1^{(m)},\dots,g_{N_c}^{(m)})$.
    \end{enumerate}

  \item \textbf{Quasi-Newton step (L-BFGS-B update):}
    \begin{enumerate}
    \item Use the past history of $(\boldsymbol{\Omega}^{(m')},\mathbf{g}^{(m')})$
    for $m'<m$ to build a low-rank approximation of the inverse Hessian
    $H^{(m)}$ (in standard L-BFGS fashion).
    \item Compute the search direction
    \[
      \mathbf{p}^{(m)} = H^{(m)} \mathbf{g}^{(m)}.
    \]
    \item Perform a line search along $\mathbf{p}^{(m)}$ (with bound handling)
    to find a step size $\eta^{(m)}$ maximizing $\mathcal{N}$:
    \[
      \boldsymbol{\Omega}^{(m+1)}
      = \mathrm{Proj}_{[\Omega_{\min},\Omega_{\max}]^{N_c}}
      \left(\boldsymbol{\Omega}^{(m)} + \eta^{(m)}\mathbf{p}^{(m)}\right),
    \]
    where $\mathrm{Proj}$ denotes component-wise clipping.
    \item Evaluate
    $\mathcal{N}^{(m+1)} = \mathcal{N}(\boldsymbol{\Omega}^{(m+1)})$.
    \end{enumerate}

  \item \textbf{Stopping criterion:}
    \begin{itemize}
      \item If $\|\mathbf{g}^{(m)}\|$ and
      $|\mathcal{N}^{(m+1)}-\mathcal{N}^{(m)}|$ fall below given tolerances,
      terminate.
    \end{itemize}
  \end{enumerate}
\end{enumerate}

\medskip
\textbf{Output:} optimized control vector $\boldsymbol{\Omega}^\star$ and
pulse $\Omega^\star(t)$ returned by the L-BFGS-B quasi-Newton procedure.
\end{minipage}
}
\caption{Pseudocode for L-BFGS-B optimal control applied to the maximization
of total non-Markovianity with bound-constrained pulse amplitudes.}
\label{alg:lbfgsb}
\end{figure*}

For completeness, we summarize the hyper-parameters values used in PPO and SAC algorithms in table \ref{tab:RL_parameters_full_switched}. 

\begin{table}[h]
\centering
\caption{Key hyperparameters of the reinforcement learning algorithms used for non-Markovian control.}
\label{tab:RL_parameters_full_switched}
\resizebox{0.7\columnwidth}{!}{%
\begin{tabular}{lcc}
\hline
\textbf{Parameter} & \textbf{SAC} & \textbf{PPO} \\
\hline
Hidden layers & [256,\,256] & [64,\,64] \\
Learning rate & $3\times10^{-4}$ & $6\times10^{-4}$ \\
Batch size & 256 & 64 \\
Buffer size & 300,000 & -- \\
Total training steps & $10^{6}$ & $10^{6}$ \\
Action range & [-5,5] & [-5,5] \\
\hline
\end{tabular}%
}
\end{table}

\begin{figure*}[h]
\centering
\fbox{%
\begin{minipage}{0.95\textwidth}
\textbf{Algorithm 3: PPO}
\medskip

\textbf{Input:} initial policy parameters $\theta$, value-function parameters $\psi$,
control bounds $[\Omega_{\min},\Omega_{\max}]$, episode length $N_t$,
hyperparameters $(\gamma_{\rm RL},\lambda_{\rm GAE},\epsilon)$.
\medskip

\begin{enumerate}
\item \textbf{Initialize} policy $\pi_\theta(a|s)$ and value network $V_\psi(s)$.

\item \textbf{Repeat} (for episodes $e=1,2,\dots$):
  \begin{enumerate}
  \item Reset environment:
    \begin{itemize}
      \item Set $\rho_1(0)=|1\rangle\!\langle1|$, $\rho_2(0)=|0\rangle\!\langle0|$.
      \item Choose initial $\Omega_0$ within bounds.
      \item Construct initial state $s_0=(0,D_0,\dot D_{-1},\Omega_0)$.
    \end{itemize}
  \item \textbf{Rollout generation:}
    \begin{enumerate}
      \item For $k=0,\dots,N_t-1$:
        \begin{enumerate}
        \item Sample action $a_k \sim \pi_\theta(\cdot|s_k)$.
        \item Update control:
        \[
           \Omega_{k+1} = \mathrm{clip}(\Omega_k + a_k,\Omega_{\min},\Omega_{\max}).
        \]
        \item Propagate $\rho_1,\rho_2$ from $t_k$ to $t_{k+1}$ under $\Omega_{k+1}$.
        \item Compute $D_{k+1}$, $\dot D_{k+1}$ and reward
        \[
           r_k = \max(0,\dot D_{k+1})
                  - \alpha(\Delta\Omega_k)^2 - \beta\Omega_{k+1}^2.
        \]
        \item Form next state $s_{k+1}$.
        \item Store $(s_k,a_k,r_k,s_{k+1})$ in a trajectory buffer.
        \end{enumerate}
    \end{enumerate}
  \item \textbf{Advantage estimation:}
    \begin{itemize}
      \item Using the trajectory $\{(s_k,a_k,r_k)\}$ and $V_\psi$, compute
      advantages $A_k$ with generalized advantage estimation (GAE) and
      returns $R_k$ (targets for $V_\psi$).
    \end{itemize}
  \item \textbf{Policy and value update:}
    \begin{enumerate}
      \item Define importance sampling ratio
      \[
        r_k = \frac{\pi_\theta(a_k|s_k)}{\pi_{\theta_{\text{old}}}(a_k|s_k)}.
      \]
      \item Maximize the clipped surrogate
      \[
      L(\theta) = \mathbb{E}\!\big[
      \min\big(r_k A_k,\,
      \text{clip}(r_k,1-\epsilon,1+\epsilon)A_k\big)\big]
      \]
      using gradient ascent on $\theta$.
      \item Minimize mean-squared error
      \[
      L_V(\psi) = \mathbb{E}\big[(V_\psi(s_k)-R_k)^2\big]
      \]
      using gradient descent on $\psi$.
    \end{enumerate}
  \item Set $\theta_{\text{old}} \leftarrow \theta$ and repeat.
  \end{enumerate}
\end{enumerate}

\medskip
\textbf{Output:} trained PPO policy $\pi_\theta(a|s)$ that maximizes total
non-Markovianity $\mathcal{N}$.
\end{minipage}
}
\caption{Pseudocode for PPO applied to maximizing the BLP non-Markovianity
measure.}
\label{alg:ppo}
\end{figure*}

\begin{figure*}[h]
\centering
\fbox{%
\begin{minipage}{0.95\textwidth}
\textbf{Algorithm 4: SAC }
\medskip

\textbf{Input:} policy parameters $\theta$, Q-network parameters
$(\phi_1,\phi_2)$, target-Q parameters $(\bar\phi_1,\bar\phi_2)$, entropy
temperature $\tau$, replay buffer $\mathcal{D}$, control bounds
$[\Omega_{\min},\Omega_{\max}]$, episode length $N_t$.
\medskip

\begin{enumerate}
\item \textbf{Initialize} stochastic policy $\pi_\theta(a|s)$, Q-networks
$Q_{\phi_1},Q_{\phi_2}$, and target networks
$Q_{\bar\phi_i} \leftarrow Q_{\phi_i}$.

\item \textbf{Repeat} (for episodes $e=1,2,\dots$):

  \begin{enumerate}
  \item Reset environment as in Algorithm~\ref{alg:ppo} to obtain $s_0$.

  \item \textbf{Rollout and data collection:}
    \begin{enumerate}
      \item For $k=0,\dots,N_t-1$:
        \begin{enumerate}
        \item Sample action $a_k \sim \pi_\theta(\cdot|s_k)$.
        \item Update control:
        \[
           \Omega_{k+1} = \mathrm{clip}(\Omega_k + a_k,\Omega_{\min},\Omega_{\max}).
        \]
        \item Propagate $\rho_1,\rho_2$ from $t_k$ to $t_{k+1}$, compute
        $D_{k+1}$, $\dot D_{k+1}$ and reward $r_k$ as in Algorithm~\ref{alg:ppo}.
        \item Form $s_{k+1}$ and store transition
        $(s_k,a_k,r_k,s_{k+1})$ in replay buffer $\mathcal{D}$.
        \end{enumerate}
    \end{enumerate}

  \item \textbf{Parameter updates:}
    \begin{enumerate}
      \item For several gradient steps per episode:
        \begin{enumerate}
        \item Sample a minibatch of transitions
        $(s,a,r,s') \sim \mathcal{D}$.
        \item Sample next actions $a'\sim\pi_\theta(\cdot|s')$.
        \item Compute target value
        \[
        y(r,s') = r + \gamma_{\rm RL}
        \big[\min_i Q_{\bar\phi_i}(s',a') - \tau \log\pi_\theta(a'|s')\big].
        \]
        \item Update Q-networks by minimizing
        \[
        L_Q(\phi_i) = \mathbb{E}_{\mathcal{D}}\big[(Q_{\phi_i}(s,a)-y)^2\big]
        \]
        for $i=1,2$.
        \item Update the policy by minimizing
        \[
        L_\pi(\theta) =
        \mathbb{E}_{s\sim\mathcal{D},\,a\sim\pi_\theta}
        \big[\tau\log\pi_\theta(a|s) - \min_i Q_{\phi_i}(s,a)\big].
        \]
        \item Update target networks with a soft update:
        \[
        \bar\phi_i \leftarrow \tau_{\rm target}\phi_i
         + (1-\tau_{\rm target})\bar\phi_i.
        \]
        \end{enumerate}
    \end{enumerate}
  \end{enumerate}
\end{enumerate}

\medskip
\textbf{Output:} trained SAC policy $\pi_\theta(a|s)$ that maximizes total
non-Markovianity $\mathcal{N}$ via entropy-regularized learning.
\end{minipage}
}
\caption{Pseudocode for SAC applied to maximizing the BLP non-Markovianity
measure in the driven open quantum system.}
\label{alg:sac}
\end{figure*}


\end{document}